\documentstyle[11pt]{article}

\def\ds{\displaystyle}
\newcommand{\bit}{\begin{itemize}}
\newcommand{\eit}{\end{itemize}}
\newcommand{\be}{\begin{equation}}
\newcommand{\ee}{\end{equation}}
\def\ep{{\mbox{\large e}}}
\newcommand{\NN}{{\mbox{I$\!$N}}}
\def\qa{{\bf q}_{\scriptscriptstyle A}}
\def\qb{{\bf q}_{\scriptscriptstyle B}}
 
\begin{document}
\begin{center}
{\huge Diffractive orbits in quantum billiards}

\vspace{1.0cm}

{\Large Nicolas Pavloff and Charles Schmit}
\end{center}

\vspace{0.5 cm}

\noindent Division de Physique Th\'eorique\footnote { Unit\'e de
Recherche des Universit\'es Paris XI et Paris VI associ\'ee au CNRS.},
Institut de Physique Nucl\'eaire,
F-91406 Orsay Cedex, France \break

\vspace{0.5 cm}
\begin{center}
{\bf abstract}
\end{center}
 We study diffractive effects in two dimensional polygonal billiards. We derive
an analytical trace formula accounting for the role of non-classical
diffractive orbits in the quantum spectrum. As an illustration the method is
applied to a triangular billiard.

\vspace{5cm}
\noindent PACS numbers :\hfill\break
\noindent 03.65.Sq Semiclassical theories and applications.\hfill\break

\noindent IPNO/TH 95-14 \hspace{0.5cm} {\it to appear in Physical Review
Letters }\hfill\break
\newpage

	During the last decade several methods based on periodic orbit (PO)
theory have been successfully employed to study quantum systems whose classical
equivalent is chaotic (see {\it e.g.} \cite{pot}). PO theory applies also when
the system is not fully hyperbolic (when some orbits appear in families
\cite{Creagh}) or integrable \cite{BT}. More recently it has been refined to
include complex orbits \cite{complex} and diffractive effects
\cite{Vat94,Whe94}. In this line we aim at studying the problem of wedge
diffraction as an extension of the standard PO theory. This is one of the
oldest and simplest example of diffraction (see {\it e.g.} \cite{Som}) and it
is also the case where the diffractive corrections to semiclassics are the more
important.

	In this Letter we calculate for the first time the role of
non-classical diffractive orbits in the spectrum of two-dimensional polygonal
billiards. We derive a trace formula embodying the contribution of diffractive
PO's to the level density (Eq. (\ref{e11})). This contribution is of order
$\sqrt{\hbar}$ smaller than the contribution of isolated PO's and is the next
order term in the trace formula. As an example the formalism is applied to a
triangular billiard with angles ($\pi/4$, $\pi/6$, $7\pi/12$) and one sees that
it provides a very accurate description of the Fourier transform of the
spectrum.

	We consider a quantum particle enclosed in a polygonal billiard
${\cal B}$ and we impose Dirichlet boun\-da\-ry conditions on the frontier
$\partial {\cal B}$. Hence the associated Green function is solution of the
following equation :

\begin{eqnarray}\label{e1}
(\Delta_{\! \scriptscriptstyle B} + k^2) G(\qb ,\qa,k) & = & \delta (\qb - \qa)
\qquad \hbox{inside} \quad {\cal B} \; , \\
G(\qb ,\qa,k) & = & 0 \qquad \hbox{on} \quad \partial{\cal B} \nonumber \; .
\end{eqnarray}

\noindent where ${\bf q}$ is a coordinate in configuration space.

	The semiclassical approximation for $G$ reads (see {\it e.g.}
\cite{Gu90})

\be\label{e2} G_0 (\qb , \qa, k) = \sum_{\qa\to \qb}
{\ep^{\ds i(k L-\mu\pi /2)}\over
\ds i \sqrt{\ds 8i\pi k L} } \; ,
\ee

\noindent where the sum is taken over all classical trajectories going from
$\qa$ to $\qb$. In (\ref{e2}) $L$ is the length of the trajectory and $\mu$ is
the associated Maslov index \cite{Gu90}. In polygonal enclosures
the boundary has no focussing components, there are no caustics and $\mu$ is
simply twice the number of bounces of the trajectory on $\partial {\cal B}$.

	In polygonal billiards the hamiltonian flow is discontinuous on the
vertices \cite{Dag93} and when the angle at a vertex is not of the form $\pi/
n$ ($n \in \NN^*$) this causes diffraction (see {\it e.g.} \cite{James}). Then,
following Keller's geometrical theory of diffraction \cite{Keller}, one is lead
to consider non-classical contributions to the Green function which are
``diffractive orbits" starting at $\qa$, going to a vertex ${\bf q}_1$ and then
to $\qb$. These orbits are non classical because at ${\bf q}_1$ the reflection
is not specular. Far from the region of discontinuity of the hamiltonian flow
the corresponding Green function may be taken to be :

\be\label{e3} G_1 (\qb , \qa, k) = G_0 ({\bf q}_1 , \qa,k)
{\cal D}_1 (\theta, \theta') G_0 (\qb , {\bf q}_1, k) \; ,
\ee

\noindent where ${\cal D}_1 (\theta, \theta')$ is a diffraction coefficient
evaluated in the solvable case of two semi infinite straight lines meeting with
an angle $\gamma$ equal to the interior angle of the polygon at ${\bf q}_1$.
$\theta$ (resp. $\theta'$) is the angle of the incoming (resp. outcoming)
trajectory at $q_1$ with the boundary. ${\cal D}_1 (\theta, \theta')$ reads
\cite{James,Keller,Pauli} :

\be\label{e4} {\cal D}_1 (\theta, \theta') = - {\ds 4\over \ds N}
{
\ds \sin(\pi /N) \sin (\theta /N) \sin (\theta' /N)
\over
\ds (\cos{\ds \pi\over N} - \cos{\ds \theta +\theta'\over\ds N} )
    (\cos{\ds \pi\over N} - \cos{\ds \theta -\theta'\over\ds N} )
}
\; , \ee

\noindent where $N = \gamma/\pi$ is not assumed to be an integer.

	As stated above one sees in expression (\ref{e4}) that when $\gamma$ is
of the form $\pi /n$, ${\cal D}_1$ is zero and there is no diffraction. Indeed
in this case a trajectory passing by ${\bf q}_1$ is the limit of a trajectory
bouncing specularly $n$ times near the vertex and a contribution of type
(\ref{e2}) accounts for the effect of the wedge. This is to be related to the
fact that in this case there exists a $n^{th}$ iterate of the flow which
is continuous \cite{Dag93}. Note also that ${\cal D}_1$ is zero if $\theta$ or
$\theta'$ is equal to 0 or $\gamma$ ({\it i.e.} in the case of a diffractive
trajectory having a segment lying on a face).

	For an orbit with several diffractive reflections at points ${\bf q}_1
... {\bf q}_\nu$ formula (\ref{e3}) becomes

\be\label{e5}
G_\nu (\qb , \qa, k) = G_0 ({\bf q}_1 , \qa, k)
\left\{ \prod_{j=1}^{\nu-1}{\cal D}_j G_0({\bf q}_{j+1} ,{\bf q}_j, k) \right\}
{\cal D}_\nu G_0(\qb , {\bf q}_\nu , k)
\; , \ee

\noindent where ${\cal D}_j$ is the diffraction coefficient at point ${\bf
q}_j$ as given by (\ref{e4}).

	In (\ref{e2}), (\ref{e3}) and (\ref{e5}) the indices 0, 1 or $\nu$ of
the Green function recall that diffractive effects are subdominant (by a factor
of order $k^{-\nu/2}$). There might be less severe non analyticities on the
boundary leading to higher order diffractive corrections. Note also that we are
using here a simple approximation for the Green function which is not valid
when the angles $\theta$ and $\theta'$ at an edge are such that the diffractive
orbit is close to be real ; in this case the coefficient ${\cal D}_1
(\theta,\theta')$ diverges. This occurs in vicinity of the line of
discontinuity of the hamiltonian flow. In order to have a formula valid in all
regions of space one should use a uniform approximation such as first provided
by Pauli \cite{Pauli} and whose general form is given in \cite{James} (see also
\cite{nous}).

	The level density $\rho (k)$ is then obtained from the Green function
by the usual formula :

\be\label{e6} \rho (k) = - {\ds 2k\over \pi} \; \mbox{Im}
\int_{\cal B} \! d^2q \; G({\bf q}, {\bf q}, k) \; . \ee

	$\rho (k)$ can be separated in a smooth function of $k$, $\bar\rho(k)$
plus an oscillating part $\widetilde\rho(k)$. The zero length trajectories in
(\ref{e6}) contribute to $\bar\rho$ and will not be considered in detail here
(see \cite{BH}). When $G$ is replaced by its semiclassical approximation
(\ref{e2}) a stationary phase evaluation of (\ref{e6}) corresponds in
considering only the contribution of classical PO's to $\widetilde\rho$. When
diffractive orbits such as (\ref{e5}) are taken into account one is lead to
consider also ``diffractive PO's" which are PO's with one or several
diffractive reflections (example of such orbits are given on Fig. 1).

	Let us consider first the contribution of classical PO's. In a
polygonal enclosure there is a drastic difference between PO's with even and
odd number of bounces. The latter ones do not remain periodic when a point of
reflection is translated along a face (they period-double into a PO with twice
as many bounces). This can be understood by remembering that, for the
phase-space coordinates transverse to the direction of an orbit, a bounce on a
straight segment leads to an inversion. On the other hand, PO's with an even
number of bounces form families which correspond to local translation parallel
to the faces of the polygon. They are neutral (or direct parabolic, see
\cite{Gu90}) PO's to which the usual trace formula does not apply ; we use a
generalization of Gutzwiller theory which is valid for the case of degenerate
PO's \cite{Creagh}. We quote here the result and leave detailed discussion for
the future \cite{nous}. A family of orbits contributes to $\widetilde \rho (k)$
as :

\be\label{e7} \widetilde \rho(k) \leftarrow \sqrt{{\ds k L\over\ds 2r\pi^3}}
\; d_\bot \cos(k r L-\pi /4) \; . \ee

	(\ref{e7}) is written for the general case of the $r^{th}$ iterate of a
primitive orbit of length $L$ ($r\in \NN^*$). $d_\bot$ is the length occupied
by the family perpendicular to the orbit's direction. It is equal to $d
\cos\phi$, where $d$ is the length occupied by the family on a face and $\phi$
the angle between the direction of the orbit and the normal to this face.

	For an isolated PO with an odd number of bounces one has the following
contribution :

\be\label{e8} \widetilde\rho(k) \leftarrow - {L\over\ds 2\pi} \cos(k r L)
\; . \ee

	Formula (\ref{e8}) holds when the number of repetitions is odd. When
$r$ is even, the $r^{th}$ iterate of an isolated orbit leads to a family and
formula (\ref{e7}) applies.

	The derivation of the contribution of a diffractive PO is patterned on
what is done in Gutzwiller's trace formula for an isolated PO. The length of a
closed diffractive orbit in the vicinity of the diffractive PO is expanded up
to second order and the trace of the Green function is evaluated by a
stationary phase approximation. The final contribution of a generic diffractive
PO with $\nu$ diffractive reflections to the oscillating part of the level
density reads :

\be\label{e11}
\widetilde{\rho} (k) \leftarrow
{\ds L\over\ds\pi} \left\{ \prod_{j=1}^{\nu} {\ds {\cal D}_j\over\ds\sqrt{\ds
8\pi k L_j}} \right\} \cos (k L -\mu\pi/2-3\nu\pi/4) \; .\ee

	In (\ref{e11}) $L_1 ... L_\nu$ are the lengths along the orbit between
two diffractive reflections. $L_1+...+L_\nu=L$ is the total length of the
diffractive PO. $\mu$ is the Maslov index which is here twice the number of
specular reflections. Formula (\ref{e11}) is the most important result of this
paper. Note that different diffractive orbits may combine if they have
diffraction points in common. Hence repetitions of a primitive diffractive
orbit appear as a special case of (\ref{e11}) ; in this case however, in the
first factor $L/\pi$ of the r.h.s. of (\ref{e11}), $L$ should be understood as
the primitive length of the orbit. The above formulae show that the
contribution of a family of orbits in of order ${\cal O}(k^{1/2})$, for an
isolated orbit it is ${\cal O}(1)$ and for a diffractive PO it is ${\cal
O}(k^{-\nu/2}$). Nevertheless we will see in the following that diffractive
orbits have a very noticeable contribution to the level density.

	We will now illustrate our approach by studying a specific example. Let
us consider a triangle with angles ($\pi/4,\pi/6,7\pi/12$). As explained above
diffraction occurs only at the vertex with angle $7\pi/12$. The scale of
lengths and wave-vectors is fixed by the value $h$ of the height going from
this vertex to the opposite face. We take $h=1$ in the following. The shortest
classical and diffractive PO's in this triangle are shown on Fig. 1.
Diffractive reflections are indicated with a black spot. Note that the first
orbits are diffractive, classical orbits (isolated or in families) occur at
greater lengths. The spectrum was computed numerically by expanding the wave
function around the vertices with angles $\pi/4$ and $\pi/6$ in ``partial
waves" which are Bessel functions with a sinusoidal dependance on the angle
defined near the vertex considered. More precisely if $r_n$ and $\varphi_n$ are
polar coordinates defined near the vertex $\pi/n$ ($n = 4$ or 6) the partial
waves in this region are of the form $J_{n m}(k r_n) \sin (n m \varphi_n)$ with
$m \in \NN^*$. One then imposes matching of the wave function and of its first
derivative along the height $h$ (see details in \cite{nous}). We determined the
first 957 levels, up to $k_{max} \simeq 96$. The accuracy of the computation
was tested by varying the number of matching points and of partial waves. We
evaluate the typical error on an eigenvalue as being of order of a hundredth of
the mean level spacing.

	In order to visualize the importance of classical and diffractive PO's
of successive lengths in the spectrum we study the regularized Fourier
transform of the level density :

\be\label{e12} F(L)=\int_0^{k_{max}}\! k \; \ep^{\ds i k L-\alpha k^2}
\rho(k) dk \; . \ee

	If $k_{max}\to +\infty$ and if the regularizing coefficient $\alpha$ is
set to zero in (\ref{e12}) $F(L)$ is just a series of delta peaks centered on
the lengths of the classical and diffractive PO's. The multiplicative factor
$k$ in (\ref{e12}) is meant to cancel the singularity $k^{-\nu/2}$ of the
contribution of a diffractive PO of type (\ref{e11}) with up to $\nu=2$
diffractive reflections. We take here $\alpha = 9/k_{max}^2$ and plot $|F(L)|$
on Fig. 2. The numerical result is represented by a thin line and the
semiclassical approach (\ref{e7},\ref{e8}) corrected by diffractive PO's
(\ref{e11}) by a thick line. We also included the contribution of $\bar\rho(k)$
in order to reproduce the initial peak at $L=0$. We see that the agreement is
excellent. Note that the existence of diffractive PO's is of great importance
for reproducing all the peaks in $|F(L)|$. This is illustrated on the figure
where their contribution (\ref{e11}) has been dashed.

	Here several comments are in order. Note first that the diffractive
PO's labelled 2 and 4 on Fig. 1 have not been included because their
diffraction coefficient is zero. Also the orbit labelled 7 on Fig. 1 has a non
standard contribution : it is an isolated orbit which accounts for boundary
effects on the family with the same length (labelled 6 on Fig. 1). In addition
to the orbits of this family it has an extra reflection on the bottom face (the
same type of orbit was considered in Ref. \cite{Lau91,Sie93}). The weight of PO
number 7 is reduced by a factor 1/2 compared to (\ref{e8}) since one integrates
only over closed orbits on one side of this limiting PO. Also we included
repetitions of diffractive PO's number 1 and 3 and they can be seen to have
still a noticeable contribution. We did not include the diffractive PO composed
by the sum of orbit 1 and 3 although it can be considered as a small
diffractive correction to the contribution of family 6. Indeed the orbit ``1+3"
lies just on the region separating real orbits from diffractive ones and as
mentioned above it can not be accounted for by a simple diffraction coefficient
such as (\ref{e4}). This type of corrections will be treated in a forthcoming
publication \cite{nous}.

	To summarize let us emphasize the important role of non classical
orbits in the spectrum of quantum billiards. The existence of these orbits
affects qualitatively the Fourier transform of the spectrum. The above example
is only one among others were the discontinuity of the classical dynamics is
linked to strong diffractive corrections to semiclassics. It was argued in
\cite{Pav93} that the same type of corrections should be taken into account for
the 3 dimensional icosahedral billiard. We expect also diffractive effects --
of the same order as those described here -- in more general billiards with
cusps (non-polygonal or with an additional external field) ; in these cases a
simple generalization of formula (\ref{e11}) accounts for the role of
diffractive PO's. We note finally that the present work illustrates that
semiclassical methods provide a very appealing tool which, when corrected with
tunneling or diffractive effects, allows to describe accurately the solution of
partial differential equations using simple geometrical methods.

	It is a pleasure to thank E. Bogomolny and D. Ullmo for fruitful
discussions. Division de Physique Th\'eorique is a Unit\'e de Recherche des
Universit\'es de Paris XI et Paris VI associ\'ee au CNRS.

\newpage
\vspace{1cm}
{\center {\Large {\bf Figure captions }}}
\vspace{1cm}

{\bf Figure 1.} ~The shortest classical and diffractive PO's in the triangle
($\pi/4,\pi/6, 7\pi/12$). All these orbits are self-retracing. For diffractive
PO's the diffraction point is marked with a black spot. Orbits 6 and 10 form
families, 5 and 7 are isolated. The lengths are given in unit of the height of
the triangle.

\vspace{1cm}

{\bf Figure 2.} ~$|F(L)|$ as a function of $L$. The thin line is the numerical
result and the thick line the semiclassical approximation (\ref{e7},\ref{e8})
with diffractive corrections (\ref{e11}). The two curves are hardly
distinguishable. The contribution of the diffractive PO's has been dashed.

\end{document}